\documentclass[9pt,twocolumn,twoside]{osajnl}

\usepackage{epsfig}
\usepackage{epstopdf}
\usepackage{verbatim}

\newcommand{\be}{\begin{equation}}
\newcommand{\ee}{\end{equation}}

\newcommand{\myeq}[1]{Equation~(\ref{eq:#1})}

\journal{ao} 

\setboolean{shortarticle}{false} 

\ifthenelse{\boolean{shortarticle}}{\colorlet{color2}{color2b}}{\colorlet{color2}{color2}} 

\title{Switchable Virtual, Augmented, and Mixed Reality through Optical Cloaking}

\author[1,*]{Joseph S. Choi}

\affil[1]{The Institute of Optics, University of Rochester, Rochester, New York 14627, USA}
\affil[*]{joecmama@gmail.com}

\dates{Compiled \today}

\ociscodes{(090.1995) Digital holography; (110.1758) Computational imaging; (230.3205) Invisibility cloaks; (100.2000) Digital image processing.}

\doi{\url{http://dx.doi.org/10.1364/ao.XX.XXXXXX}}

\begin{abstract}
A switchable virtual reality (VR), augmented reality (AR), and mixed reality (MR) system is proposed using digital optical cloaking.  Optical cloaking allows completely opaque VR devices to be ``cloaked,'' switching to AR or MR while providing correct three-dimensional (3D) parallax and perspective of the real world, without the need for transparent optics.  
On the other hand, 3D capture and display devices with non-zero thicknesses, require optical cloaking to properly display captured reality. 
A simplified stereoscopic system with two cameras and existing VR systems can be an approximation for limited VR, AR, or MR.  To provide true 3D visual effects, multiple input cameras, a 3D display, and a simple linear calculation amounting to cloaking can be used. 
Since the display size requirements for VR, AR, and MR are usually small, with increasing computing power and pixel densities, the framework presented here can provide a widely deployable VR, AR, MR design. 
\end{abstract}

\setboolean{displaycopyright}{true}

\begin{document}

\maketitle
\thispagestyle{fancy}

\ifthenelse{\boolean{shortarticle}}{\ifthenelse{\boolean{singlecolumn}}{\abscontentformatted}{\abscontent}}{}

\section{Introduction}
``Virtual Reality'' (VR), ``Augmented Reality'' (AR), ``Mixed Reality'' (MR) incorporate simulated worlds with or without some portion of the real world of the user.
VR presents a virtual world to the user that is not identical to the real world; in AR, simulated objects are overlaid (`augmented') onto the real world; and in MR the virtual or simulated world is `mixed' with the real world to present a combined world~\cite{Kelly-MagicLeap-May2016}.  
For this work, untouched content of reality (such as what our eyes observe without any devices) will be called `real reality,' while simulated content that is different from real reality, such as content in VR, AR, and MR will be called `simulated reality.'
Many of the current devices that incorporate simulated reality are head-mounted displays with a history of designs and applications that include the military, commercial, and consumer space~\cite{Cakmakci-Rolland-2006,Kress-2013}.

VR devices typically display two stereoscopic images, one image to each eye, at a time.  
Most, if not all, VR devices do not allow users to see through the device as they are opaque.  However, some use detection schemes to then add on some level of simulated images of the environment of the user.  Also, lenses are usually placed between the images and eyes to allow the eyes to focus on the images with ease (as if they appear from ``infinity'')~\cite{Google-cardboard-2015}.
AR devices can display simulated graphics onto glasses worn by a user~\cite{Kress-2013}, or add simulated content onto a camera generated display screen. 
MR allows users to interact with the simulated content that is incorporated in with the real world, with the real world (or portions of it) still being visible.
Many AR or MR devices incorporate the reality of the user through transparent devices, so the user can see the real environment directly.

Let us first discuss an important concept in describing light for image capturing and display purposes.
Light is an electromagnetic wave that oscillates in time, or a combination of multiple waves.  At a given point in time, each `component' of light can be described by its color (frequency), amount/strength (irradiance), position, direction, and phase.
Since human eyes cannot observe the `phase' of visible light waves, ignoring phase allows us to describe light as rays and simplifies the calculation of light propagation~\cite{FG-GO-2004,Choi-Cloak2-2015,Born-Wolf-2010}.
In particular, for a given time, a light ray can be described by its position and direction, together with its color and irradiance.  This is called a `light field' and collection of light fields can allow for 3D images to be captured and displayed~\cite{Lam-2015}.  To simplify our discussion in this work, we will assume that the color and irradiance of a light field is known (which can be measured with detectors such as those in cameras), and simply use its direction and position to describe the light field.

Some research and development of 3D light capturing and display have looked into capturing and using light fields, together with computational imaging and post-processing calculations and effects~\cite{Wetzstein-course-2012}.  Light field ideas predate many of the recent research and developments in this field, but with increased computational power and engineering, the quality and capabilities have become more practical than before.  
Even groups working in the VR, AR, or MR fields have begun to investigate using light fields~\cite{Kelly-MagicLeap-May2016}.

For 3D capture and display devices, including those used in AR or MR, if the cameras or sensors that capture images are located at different positions than where the image display pixels are located, optical ray propagation is necessary.  Otherwise, correct 3D perspective will not be observed by the user, particularly when these devices begin to generate sufficiently high resolution and quality in displayed images.  The correct optical ray propagation is to essentially ``cloak'' the space between the cameras/sensors and the display pixels, and can be applied to digital images in a simple manner~\cite{Choi-Cloak3-2016}.
Some researchers have begun investigating optical cloaking in these simulated devices~\cite{Choi-patent-2016,Howlett-2017}.
In this work, a few methods are proposed to combine (or intentionally not combine) real reality with simulated reality through the use of optical cloaking, geared for digital image capture, processing, and display.  This enables a switchable VR, AR, and MR system that does not require transparent glasses or devices, and can generalize to 3D imaging and display techniques.

\section{Simplified Model}
\begin{figure}[htbp]
\centering
\fbox{\includegraphics[scale=1.0]{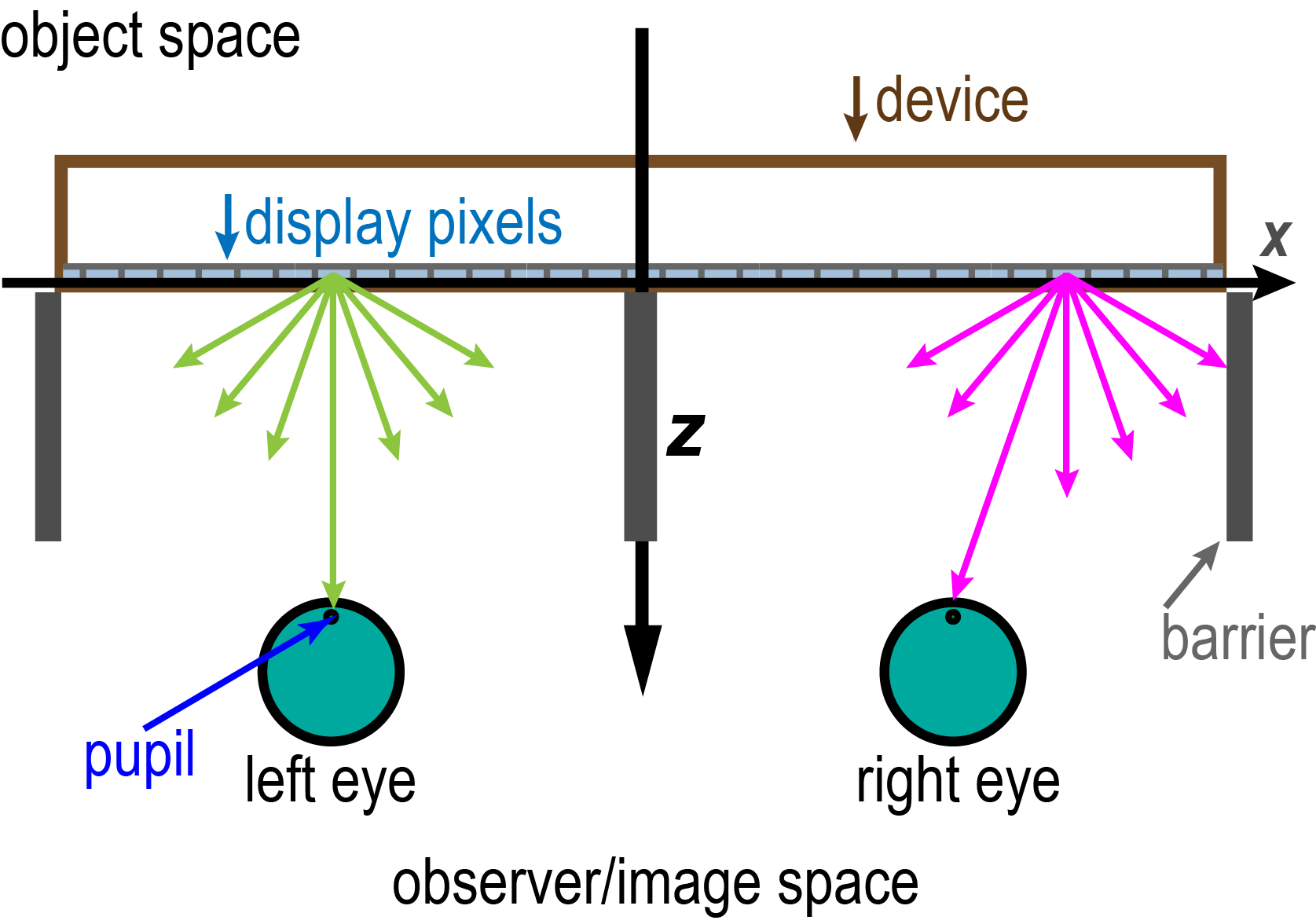}}
\caption{
\textbf{VR for point-like pupil and point-like pixel approximation.}  
For this extremely simplified model, only a single ray (the longer arrow) from each pixel enters the pupil.  Thus, a flat display screen that emits rays of light in all directions in front of it, still acts like a `light field' display.
Barriers can be used to block unwanted rays, if desired.
(A few example rays are drawn.)
}
\label{fig:VR_AR_1_VR-1}
\end{figure}

Let's first assume that the pupil of the human eye is point-like in spatial extent, and that each display pixel generates light from a single point in space.
These assumptions are not correct and are relaxed later, but will be used initially to simplify explaining some initial concepts.
With these simplifications, even a flat two-dimensional (2D) image from a display screen acts as a light field display.
This is because only one direction of light ray is emitted from each pixel into the pupil, as there is only one ray connecting these two ``points'' (see Figure~\ref{fig:VR_AR_1_VR-1}).
Thus, for small pupils and small pixels, a 2D display screen approximates a 3D light field display.  

\begin{figure}[htbp]
\centering
\fbox{\includegraphics[scale=1.0]{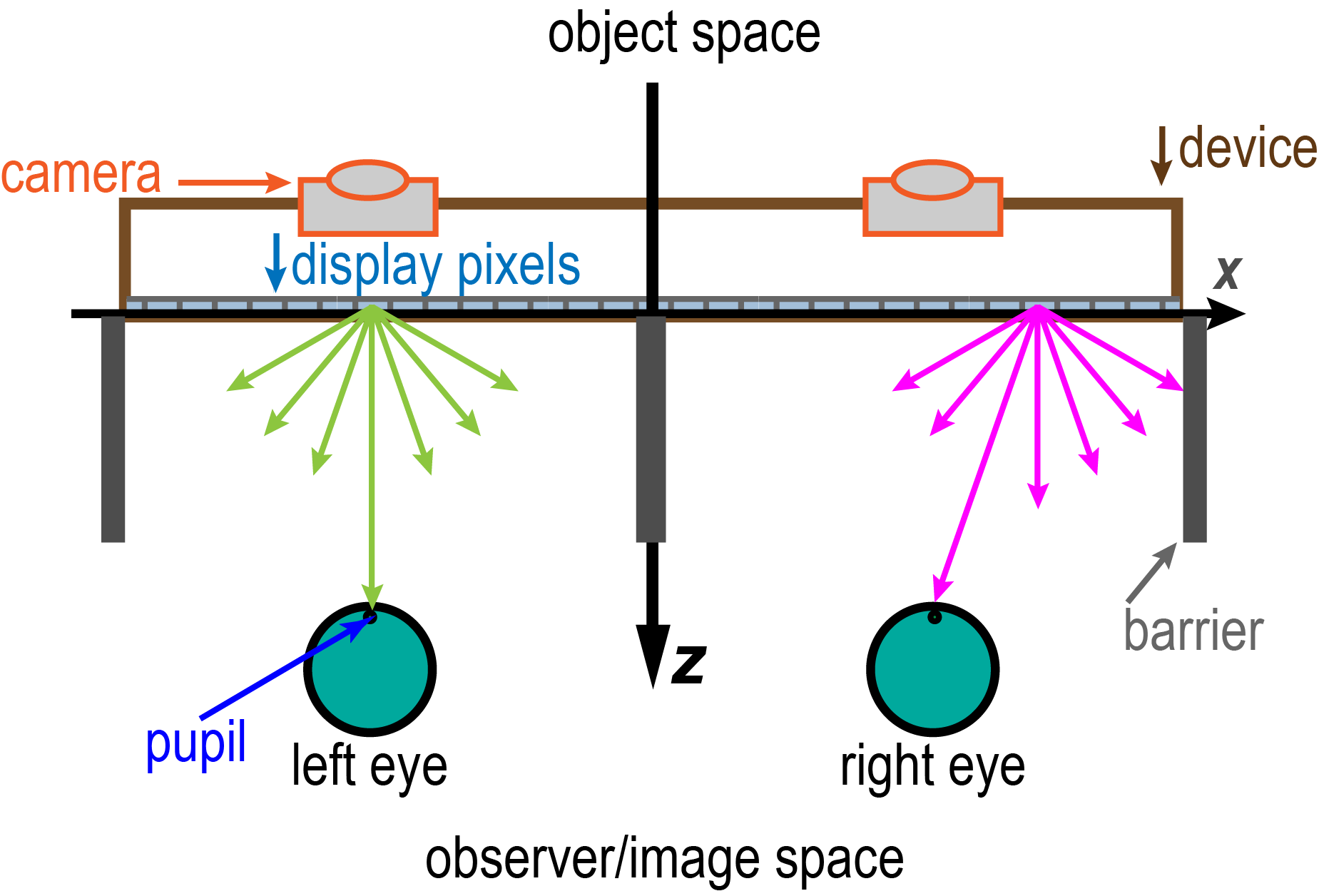}}
\caption{
\textbf{Simple switchable VR/AR/MR for point-like pupil approximation.}  
Images/videos captured by two cameras are displayed, such that the left (or right) camera image is seen by the left (or right) eye.
Physical barriers separate left and right images, and can block unwanted rays.
(A few example rays drawn.)
}
\label{fig:VR_AR_1_cam1}
\end{figure}

In a simplified VR/AR/MR switching design, two cameras could be separated by the interpupillary distance of the user's eyes.  The centers of each camera can be placed directly in front of each pupil of the eye (See Figure~\ref{fig:VR_AR_1_cam1}).  
The images or videos of each camera would be displayed on the display pixels that are seen by the eyes, the left camera image displayed for the left eye, and similarly for the right eye.

The resulting background scenery using the setup from Figure~\ref{fig:VR_AR_1_cam1} will be different than what the observer would see without any displays or devices.  The images seen are a `zoomed' view of the real scenery, since the observer roughly views the background scene as if his/her eyes are located at the camera positions, instead of his/her current pupil location.  This zooming effect can be adjusted digitally to some extent, but can be corrected or changed in real-time with a light field device~\cite{Choi-patent-3-2016}.

Figure~\ref{fig:VR_AR_1_cam1} allows for the stereoscopic capture of the scenery in front of the user of the device in real-time.  This approximated `real reality' can be captured and displayed for the user with changes in magnification or translation in space and time, for additional visual effects if desired.  This can then be combined with simulated reality using algorithms for 3D reconstruction and placement, using the stereoscopic depth information, to combine computer generated graphics, allowing for an AR or MR experience.  The device can be switched to a VR experience by only providing simulated reality, without the images captured by the cameras.  This is then a very basic, switchable VR, AR, and MR design, within the point-like pupil and pixel approximation.  An advantage of such a simple design in addition to pedagogy, is the ease of implementation, followed by the ease of users to capture and share stereoscopic images, videos, and experiences.

Although Figures~\ref{fig:VR_AR_1_VR-1} and ~\ref{fig:VR_AR_1_cam1} may have some uses for VR, AR, or MR, a problem with it is that pupils are \emph{not} points and actually have non-zero, finite size.  This allows light rays from different pixels (and different angles) to be detected by the same point (cone or rod, for example) in the retina of the eye.   
Because of this `mixing' of multiple display pixels, as sensed by the same cell in the retina, the image may appear blurred.
There are some methods to alleviate this blurring.  One method is to move the display screen farther from the eyes, so that each pixel subtends a smaller range of angles of light into the pupil, approximating conditions that are closer to the point-like pupil approximation.  For example, the display screen can first be brought close to the eyes until the brain `merges' the two stereoscopic images, and then the screen can be moved farther until the 3D image appears sharp enough.

\begin{figure}[htbp]
\centering
\fbox{\includegraphics[scale=1.0]{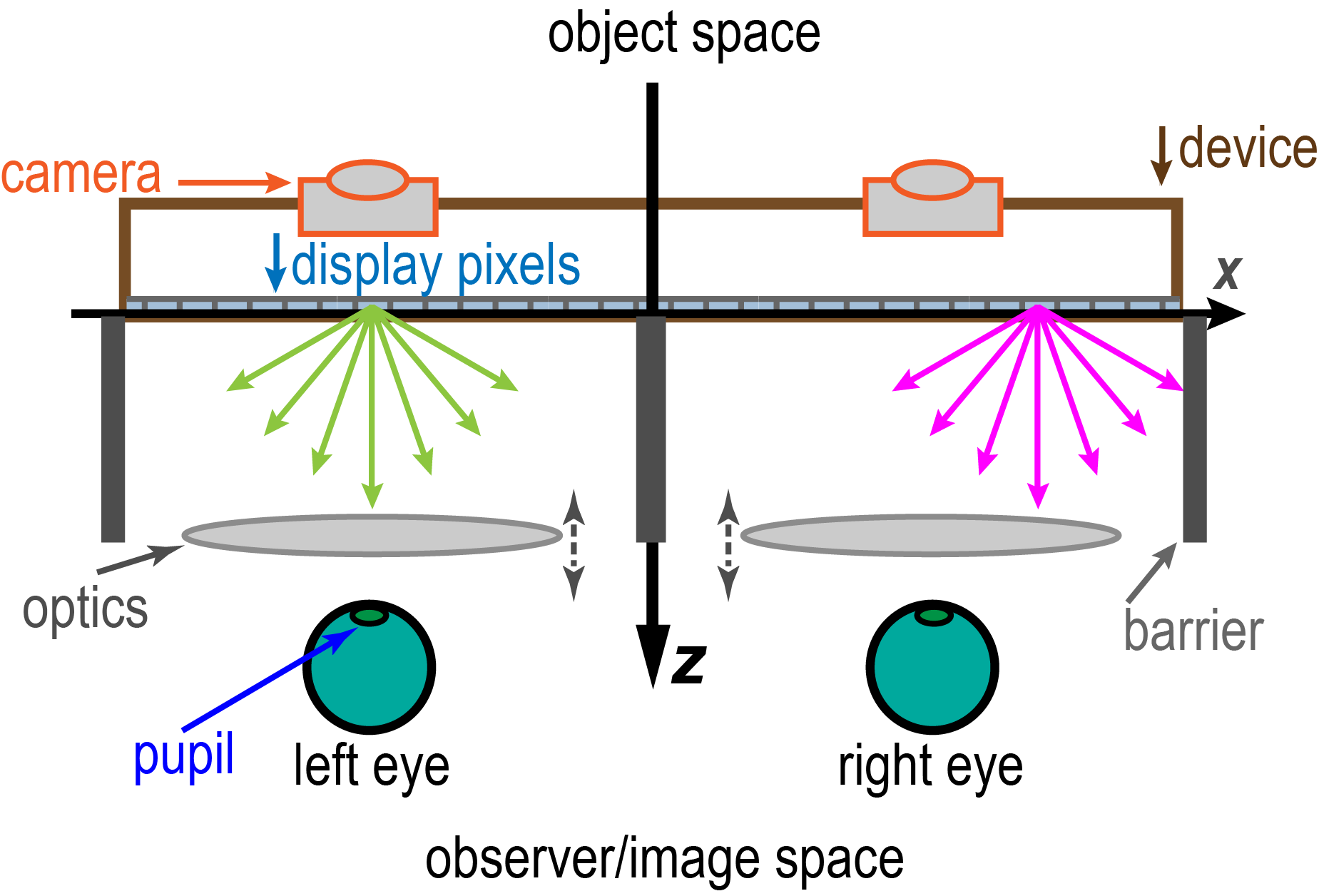}}
\caption{
\textbf{Simple switchable VR/AR/MR with optics.}  
Two cameras are used to capture stereoscopic images that are displayed to left eye or right eye.  
Optics can be used, and positions varied, to make the objects from the screen appear appropriately for the user.  
}
\label{fig:VR_AR_1_cam1-lens1}
\end{figure}

Another method to reduce the blurring is to use two lenses (or more complex optics), one for each eye, with each lens separated from the display screen by its focal length (see Figure~\ref{fig:VR_AR_1_cam1-lens1}).  This is a common method employed by VR systems~\cite{Google-cardboard-2015}.  This allows the display screen objects (or `images') to appear as if located at `infinity,' which is easier for the eye to view than otherwise. 
However, all objects on the display surface are located at or near the same position in reality, despite \emph{seeming} to appear at different depths from the eye (due to the cues given from the different left and right images displayed).  This is a typical problem with many 3D stereoscopic images, a cause of vergence-accommodation conflicts and others, due to the actual light rays that are received by the eyes being different from what the brain perceives.

\section{Optical Cloaking Principles}
For many image capture and display devices, the image capture location (for example, the locations of the camera(s) that capture the scenery ahead) is not the same as the location of the display pixels.  This is due to the thickness of the device, or simply where the camera(s) and display pixels are located relative to each other.  Because of this difference, simply projecting the images captured by the input (camera(s)) onto the output (display screen) will generate a misalignment in the position and appearance of what is displayed, compared to the real reality (what the observer would see when the device is removed)~\cite{Choi-Cloak3-2016}.  To solve for this, and to properly recreate real reality, the device must appear as if it is not present to the observer.  Hence, the device should be cloaked to be `invisible,' or transparent, by its own display.
In addition, to properly account for the non-zero, finite size of the observer pupil, light field rays should be both captured and displayed to different points of the pupil, or to different positions of the retina of the observer, providing a proper 3D image.
By utilizing 3D optical cloaking~\cite{Pendry-2006,Leonhardt-2006,Choi-Cloak1-2014,Choi-Cloak3-2016}, we can solve these two issues.  

This then allows for a proper, switchable VR and AR/MR device.  
We essentially `cloak' the VR/AR/MR device, or parts of it, to make it transparent, by collecting light rays and properly displaying them.  The captured light rays can be combined with algorithms and simulated graphics onto the display.
Such a scheme is VR if a complete virtual world is displayed only.  It can be switched to AR or MR when simulated reality is combined with the real reality, the latter being displayed by the cloaked light rays.  With high resolution and fast computational power, this then can provide an idealized switchable VR, AR, MR device.

There are many methods to capture and recreate light field rays.  Regardless of how rays are captured or displayed, one key for proper recreation of real reality for a device, is to apply cloaking calculations that make parts or all of the device transparent~\cite{Choi-Cloak1-2014,Choi-patent-1-2015,Choi-Cloak3-2016,Choi-patent-2-2016}.
The calculations for cloaking simply require that for each `input' ray of light incident on the device, an `output' ray of light should exit the device such that it appears to have traveled in a straight line from the input ray, in the same direction.  
This must be done for all rays of light to achieve omnidirectional cloaking, but can be done for some of the light rays to achieve cloaking partially in space or angles.  This allows the device to become cloaked, and hence transparent to the observer, within the design limits for which it is built.  

\begin{figure}[htbp]
\centering
\fbox{\includegraphics[scale=1.0]{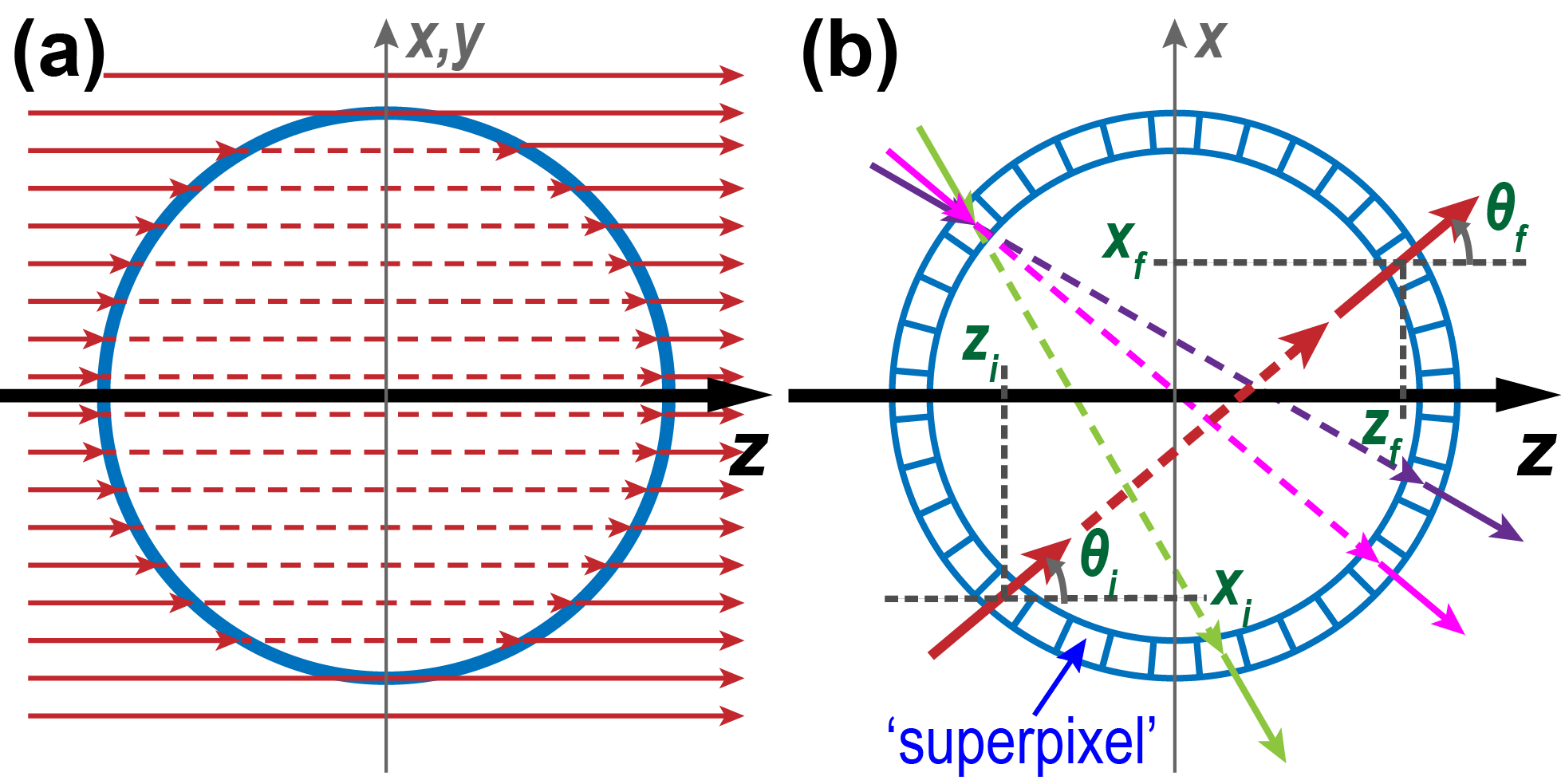}}
\caption{
\textbf{(a) Spherically symmetric cloak.}  Example rays (solid arrows) that enter and exit the cloak.  
Dashed arrows show how each ray \emph{appears} to have traveled inside the cloak/device in a straight line.
\textbf{(b) Discretized, symmetric cloak.}  
Solid arrows depict some entering and exiting light rays of the discretized surface of the device.  Each `superpixel' in space can both detect and emit multiple, discrete ray positions and angles.
(Arbitrarily shaped, and even continuous cloaks, follow same principles.)
}
\label{fig:cloak-ray-1}
\end{figure}

As an example, Figure~\ref{fig:cloak-ray-1}(a) shows some rays for a spherically shaped, or circularly shaped cloaking device, but this principle applies to arbitrary shapes, too~\cite{Choi-Cloak3-2016}.
For practical implementation, we can discretize space so that the surface is composed of `superpixels' as shown in Figure~\ref{fig:cloak-ray-1}(b).
We can quantify the mapping of input ray to output ray, though the same calculations apply for continuous space that is not discretized.
Given a transverse position $x_i$, angle $\theta_i$, and longitudinal position $z_i$ of the input ray (see Figure~\ref{fig:cloak-ray-1}(b)); the output ray has the same variable names but with subfix `$_f$'):
\be
\begin{bmatrix}
x_f \\
\tan\theta_f
\end{bmatrix}_{z=z_f}
=
\begin{bmatrix}
1 & (z_f-z_i) \\
0 & 1
\end{bmatrix}
\begin{bmatrix}
x_i \\
\tan\theta_i
\end{bmatrix}_{z=z_i}
\label{eq:Cloak-map-1}
.
\ee
We have assumed rotational symmetry about the center axis ($\bf{z}$), though the propagation through arbitrary shapes without rotational symmetry is easily derived from \myeq{Cloak-map-1}.
$L=(z_f-z_i)$ is constant for a planar cloak as shown in Figure~\ref{fig:integral-cloak-1}(b).

\section{Light field designs}
\begin{figure}[htbp]
\centering
\fbox{\includegraphics[scale=1.0]{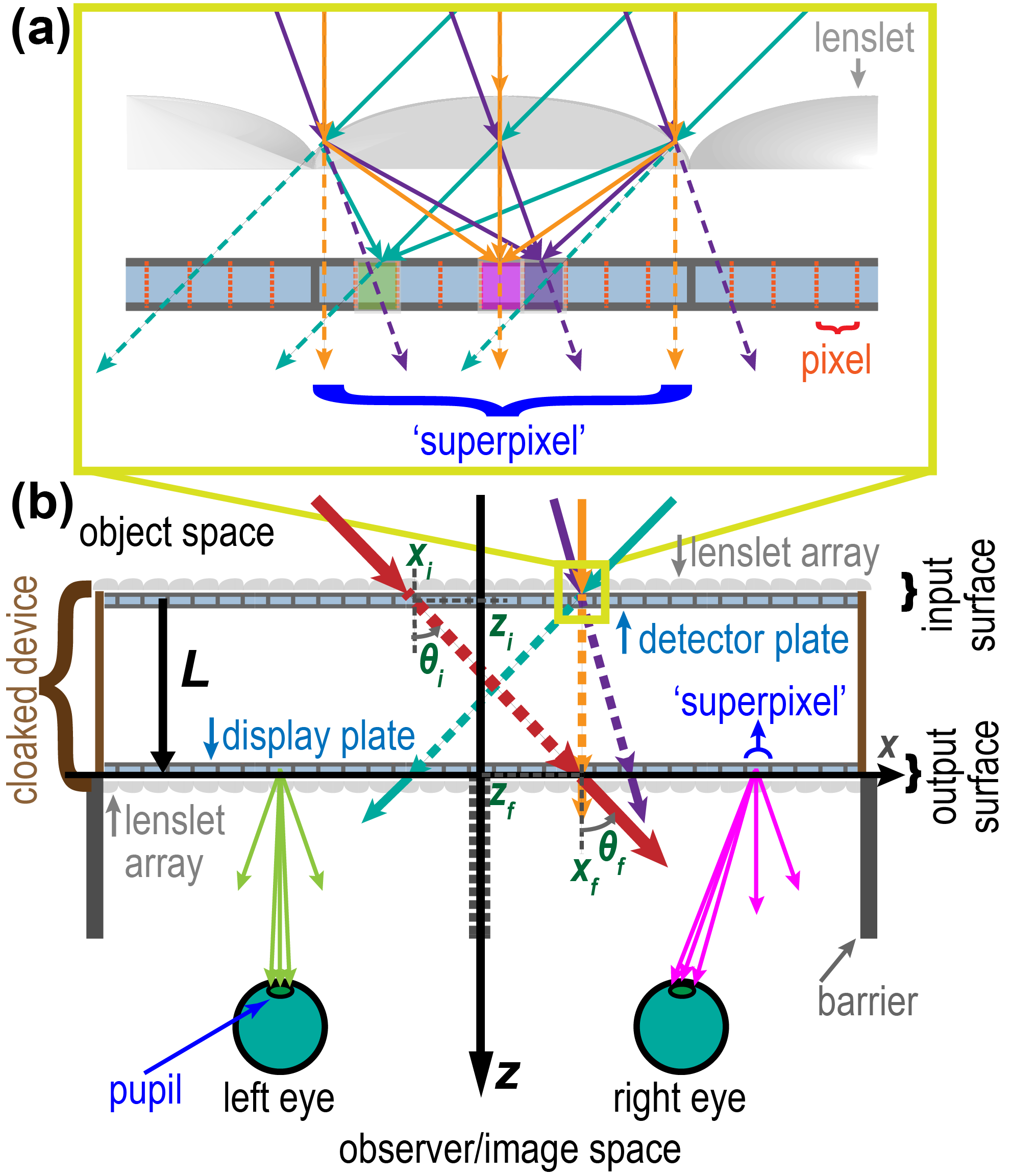}}
\caption{
\textbf{(a) Digital integral imaging detection.}  
(Zoomed out portion of \textbf{(b)}).  
A `superpixel,' placed at the focusing plane of a lenslet, collects rays with the same position as the lens.  These rays are spatially separated into smaller pixels, such that one ray angle maps to one pixel.
Display (output) is the reverse of detection (input) method shown in this figure.
\textbf{(b) A VR/AR/MR device using `digital integral cloaking.'}  
Cross-section of two parallel, 2D surfaces, with a few sample rays.  
The input `surface' (lens array + plate) captures input rays of light.  
Output surface displays light field rays as if they passed through ambient space (dashed lines).  
Output rays are designed to enter the pupils of the user.
The center barrier can be removed for a handheld device.
}
\label{fig:integral-cloak-1}
\end{figure}

Figure~\ref{fig:integral-cloak-1} shows an example of using digital integral cloaking~\cite{Choi-Cloak3-2016} for a 3D, light field, VR/AR/MR switchable device.
The design shows only the $\mathbf{x-z}$ dimension, but the principles are the same for the other $(\mathbf{y-z})$ dimension, too. 
For simplicity, the design uses input surfaces (which collect input light rays) and output surfaces (which emit/display output light rays) that are parallel and straight.  However, for arbitrary surface shapes, \myeq{Cloak-map-1} can be used so that each ray appears to propagate through the device in a straight line.
The center barrier can be used to prevent cross-talk between left and right images, if desired.

For Figure~\ref{fig:integral-cloak-1} or similar 3D light field devices, if all of the output surface is to be utilized, then the size of the input surface will have to be large enough to collect sufficient rays of light for, and near, the edges of the output surface.  For example, for the straight, parallel surfaces design shown in Figure~\ref{fig:integral-cloak-1}(b), assuming that the width covered by the output display pixels is $W_d$, then in the same $\mathbf{x-z}$ plane, the total width covered by the input pixels must be at least $W_c=(W_d + 2 \cdot L \cdot \tan(FOV_l /2))$ wide, with at least $(L \cdot \tan(FOV_l /2))$ longer width than the output width on each side.  Here, $FOV_l$ is the field-of-view of the output/display system.

The barriers can be used to also separate eyes of the user from the display screen at a fixed (or adjustable) distance, for optimized viewing.  
For Figure~\ref{fig:integral-cloak-1} some or all of the barriers can be removed or made detachable for varying effects and control, including making it a handheld device.
For a handheld device, allowing the display screen to send light fields beyond just two fixed eye pupils is ideal, so that the device can be viewed from varying angles and positions.

There are other designs that can be useful for VR, AR, and MR.  These include using various methods for input and output for 3D light field rays.  For example, light field rays can be collected and/or displayed through detectors and display pixels and lenslet arrays, microlenslet arrays~\cite{Lanman-2013}, diffractive optics, holographic elements, parallax barrier techniques~\cite{Kim-parallax-2016}, spatial light modulators, other methods, or a combination of these.

\begin{figure}[htbp]
\centering
\fbox{\includegraphics[scale=1.0]{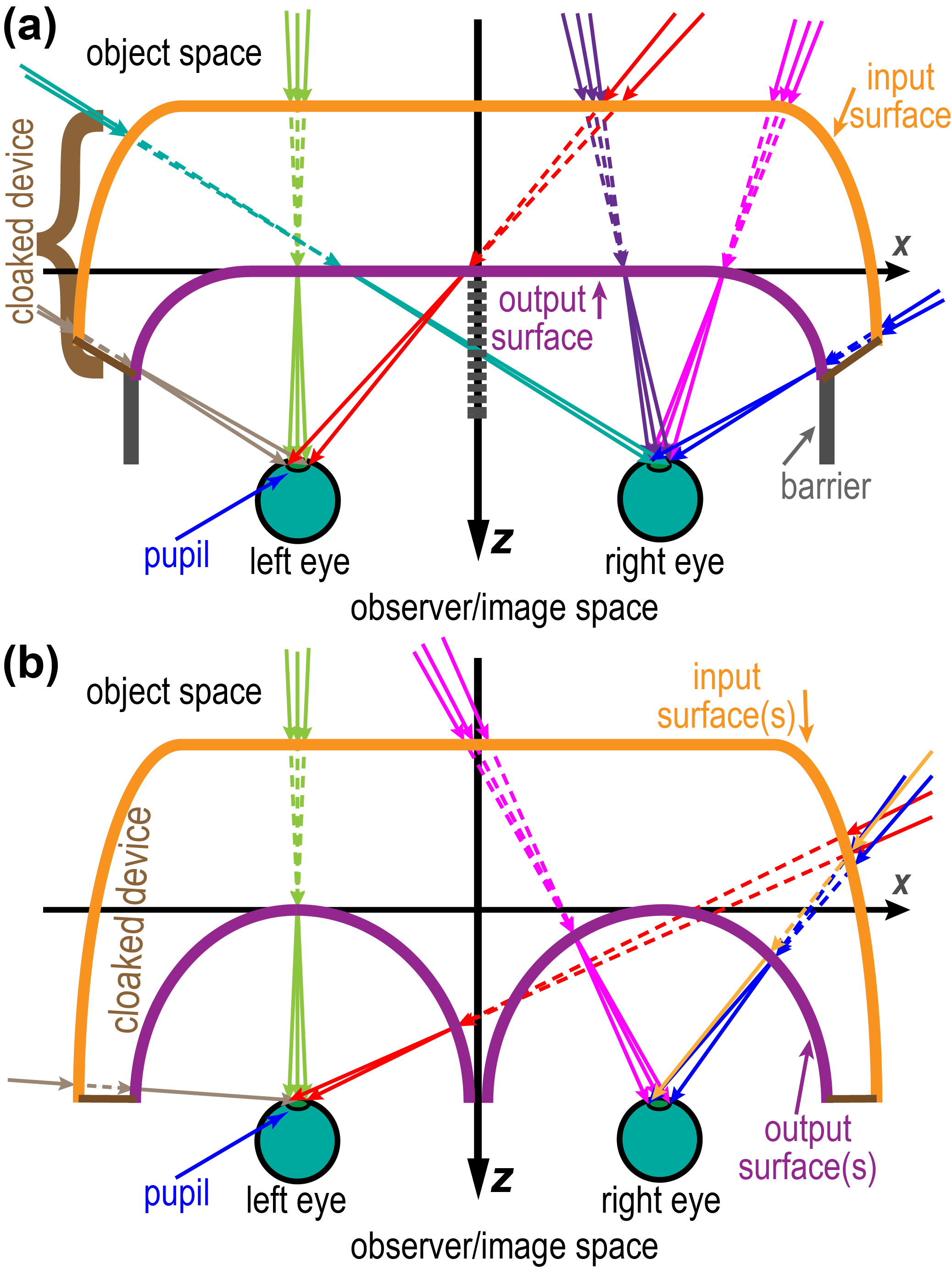}}
\caption{
\textbf{(a) Large field-of-view, switchable VR, AR, MR device using cloaking and light field.}  
Cross-section of a curved device, with some sample rays.  
The input surface captures light field rays from the object space.  
Output surface displays light field rays as if they passed through ambient space only (in straight trajectories, i.e., \myeq{Cloak-map-1}, shown in dashed lines).  
Barriers may or may not be needed, and the center barrier can be removed to allow rays to enter from left to right and vice versa.
\textbf{(b) Another curved device with large field-of-view.}  
One of many possible variations.  This design can provide 3D images with minimal cross-talk between left and right eyes.  Also, the display superpixels can have less stringent requirements for the range of light ray angles to output, compared to flatter displays.
}
\label{fig:VR_AR_1_cloak-2}
\end{figure}

It is worth discussing design optimization for light field devices to generate practical and cost-effective solutions.  When using light field rays to output into the pupils of observers, light rays that would not enter the pupils are not necessary to output.  Utilizing this fact can reduce input power requirements, pixel density or light field resolution requirements, and computational resources.  
In fact, since the foveal region, or central vision, is the sharpest for humans, the distribution of resources (pixels, lenslets, other light field components, computing power, etc.) can be focused for the central region of the output surface.
For example, on the output display, fewer ray angles can be generated for the edges of the field-of-view compared to the center.
Correspondingly, the input surface can be adjusted to collect not only those rays necessary for the display output surface, but concentrate detectors where more light rays need to be collected for the display.  
Figures~\ref{fig:VR_AR_1_cloak-2}(a) and~\ref{fig:VR_AR_1_cloak-2}(b) illustrate some of this optimization, though this can be applied to Figure~\ref{fig:integral-cloak-1} and other designs as well.
Also, to view different parts of a scenery, many users tend to move and rotate their head and body, rather than rotating or moving their eyes only.  So the user experience can still be excellent even if generating high resolution images is concentrated near the field-of-view centers, which stays fixed relative to the eyes when the user moves his/her head and body.

Figures~\ref{fig:VR_AR_1_cloak-2}(a) and~\ref{fig:VR_AR_1_cloak-2}(b) show cross-sections of other switchable VR, AR, MR designs that have curved surfaces.  One advantage of the curved surfaces is the large field-of-view display that is possible despite each superpixel being allowed to have relatively small field-of-view.  These designs can be easily extended to arbitrary surfaces, including those that do not necessarily require $\mathbf{x,y}$ symmetry.  
Output rays can be optimized for the user pupils, with increased light field rays devoted to the foveal vision, while fewer light rays are generated for the outer field-of-view.
As shown in Figure~\ref{fig:VR_AR_1_cloak-2}(b), some rays can appear to have traveled in and out of the device, to connect the output ray with its input ray in a straight trajectory to meet \myeq{Cloak-map-1}.

\section{Discussion}
Note that devices that only collect input light rays along a line on the input surface (for example, along $y=$constant on the input surface of Figures~\ref{fig:integral-cloak-1} and~\ref{fig:VR_AR_1_cloak-2}), can still be a useful light field VR, AR, or MR device.  This is because the range of pupil positions of users may not necessarily be large in the vertical ($\mathbf{y}$) direction.  Also, users' eyes can rotate (vertically or horizontally), but to generate 3D views for rotation of the eyes (assuming pupils remain in the same positions relative to the display), detecting light field rays for all possible vertical positions on the input surface may not be necessary.  Large enough field-of-views of the input detection method, combined with zooming may be sufficient~\cite{Choi-Cloak3-2016}.  Thus, some useful designs may include cylindrical lenslets (slanted or straight) for input and/or output surfaces, parallax barriers, or arrays of cameras along a line on the input surface or on a limited area of the input surface.  However, to improve the vertical and horizontal alignment and magnification, spherical lenslet arrays, ``fly's eye'' lenslet arrays, or various spatial light modulators can also be used.

The switchable VR/AR/MR designs presented, or any other 3D capture or display devices, can change the distance of propagation between the input and output locations ($(z_f-z_i)$ in \myeq{Cloak-map-1}, or $L$ in Figure~\ref{fig:integral-cloak-1}(b)) to be different than the actual physical separation, in post-processing.  This then can produce a `zooming' effect for the image displayed using such rays of light, so that the new image appears as if viewed from a location closer or farther than if the device was properly `cloaked' (where $(z_f-z_i)$ in \myeq{Cloak-map-1} matches the actual physical distance between input and output rays)~\cite{Choi-patent-3-2016}.
This propagation distance can be dynamically controlled and changed, so that the device is not limited to showing a static zoom distance, but varying zoom distances in 3D, that can be controlled by the user.

To combine simulated reality onto real reality effectively, the 3D real reality may need to be reconstructed or approximated, even if partially, in 3D space.  This can be done through computational algorithms that estimate the depth of various objects captured by the light field ray positions, angles, and input and output surfaces~\cite{Choi-patent-3-2016}.  Utilization of research in computational imaging, image interpolation, discretized vs. continuous space, computer vision, and machine learning algorithms can be helpful.  
One important issue could be the lag in time, between image capture to the display of images (simulated, real, or combination).  Concentrating sufficient computational power and speed into reducing this lag time to below what is noticeable by the observer is necessary for smooth adoption.  As computing speed and the technological capabilities of processors increase, this can be effectively minimized.

\section{Conclusion}
In this work, various methods and designs for switchable VR, AR, MR devices have been presented.  These designs that utilize optical cloaking can present images/videos in front of the user, with correct 3D perspective, parallax, and alignment, without the need for transparent display optics.  This can increase the field-of-view of the device, and may provide less stringent manufacturing requirements than using transparent hardware.
A key element for these devices is to recreate, or approximate, the light field rays that would be observed by the user without the device, then add simulated reality and other effects as desired.
Combined with methods to generate high resolution light field rays~\cite{Choi-patent-2016}, an idealized 3D image capture and display device for VR, AR, and MR, can be widely implemented.


\bibliography{VR_AR-bib}

 

\end{document}